\documentclass[superscriptaddress, nofootinbib, preprintnumbers, twocolumn]{revtex4}

\usepackage{amsmath}	
\usepackage{amsthm}		
\usepackage{amssymb}	
\usepackage{eufrak}
\usepackage{datetime}
\usepackage{graphicx}
\usepackage{xcolor}
\usepackage{verbatim}
\usepackage{bm}
\usepackage{float}
\usepackage{braket}
\usepackage[colorlinks=true, citecolor=midblue, linkcolor=midblue, urlcolor=midblue]{hyperref}

\usepackage{setspace}

\usepackage{amsmath}	
\usepackage{amsthm}		
\usepackage{amssymb}	
\usepackage{eufrak}
\usepackage{datetime}
\usepackage{graphicx}
\usepackage{xcolor}
\usepackage{verbatim}
\usepackage{bm}
\usepackage{float}
\usepackage{todonotes}

\usepackage[colorlinks=true, citecolor=midblue, linkcolor=midblue, urlcolor=midblue]{hyperref}

\usepackage{setspace}

\definecolor{grey}{rgb}{0.4,0.4,0.4}
\definecolor{dullmagenta}{rgb}{0.4,0,0.4}
\definecolor{darkblue}{rgb}{0,0,0.4}
\definecolor{midblue}{rgb}{0,0,0.5}
\definecolor{midred}{rgb}{0.5,0,0}
\definecolor{orange}{rgb}{1,0.5,0}
\definecolor{lightbrown}{rgb}{0.75,0.5,0.25}
\definecolor{tan}{cmyk}{0.14,0.42,0.56,0}
\definecolor{djunglegreen}{cmyk}{0.99,0,0.52,0}
\definecolor{lightgreen}{rgb}{0,1,0}
\definecolor{olivegreen}{cmyk}{0.64,0,0.95,0.40}
\definecolor{midgreen}{rgb}{0.0,0.675,0.0}
\definecolor{darkgreen}{rgb}{0,0.5,0}

\smallskipamount

\settimeformat{ampmtime}

\usepackage{float}

\begin{document}

\title{Testing the Number of Neutrino Species with a Global Fit of Neutrino Data}
\author{Manuel Ettengruber}
\email{manuel@mpp.mpg.de}
\affiliation{
	Arnold Sommerfeld Center,
	Ludwig-Maximilians-Universit{\"a}t,
	Theresienstra{\ss}e 37,
	80333 M{\"u}nchen,
	Germany,}
\affiliation{
	Max-Planck-Institut f{\"u}r Physik,
	F{\"o}hringer Ring 6,
	80805 M{\"u}nchen,
	Germany}
 
\author{Alan Zander}\email{alan.zander@tum.de}
\affiliation{Technical University Munich (TUM), James-Franck-Strasse 1, 85748 Garching, Germany}

\author{Philipp Eller}\email{philipp.eller@tum.de}
\affiliation{Technical University Munich (TUM), James-Franck-Strasse 1, 85748 Garching, Germany}

\date{\formatdate{\day}{\month}{\year}, \currenttime}

\begin{abstract}
We present the first experimental constraints on models with many additional neutrino species in an analysis of current neutrino data. These types of models are motivated as a solution to the hierarchy problem by lowering the species scale of gravity to TeV. Additionally, they offer a natural mechanism to generate small neutrino masses and provide interesting dark matter candidates. This study analyzes data from DayaBay, KamLAND, MINOS, NO$\nu$A and KATRIN. We do not find evidence for the presence of any additional neutrino species, therefore we report lower bounds on the allowed number of neutrino species realized in nature. For the normal/inverted neutrino mass ordering, we can give a lower bound on the number of neutrino species of $\mathcal{O}(30)$ and $\mathcal{O}(100)$, respectively, over a large range of the parameter space. 
\end{abstract}

\maketitle

\section{Introduction}
\label{sec:Introduction}
Neutrino oscillations are so far the only phenomena of known particles that cannot be explained within the Standard Model (SM). The usual approach to explain them is by giving the neutrino a mass. Neutrino oscillation would then result from a mismatch between the flavor and the mass basis. Despite the success of this framework, several questions remain unanswered like the absolute value of neutrino masses and what mechanism generates them.
\par 
Of course, it could be the usual Higgs-Mechanism with a very small Yukawa coupling. But first, one could be puzzled about the smallness of the Yukawa coupling and secondly, the fact that the right-handed partner,$\ \nu_R$, would be uncharged under the SM gauge group opens the gate for very different mechanisms that would be a candidate for mass generation for the neutrino. For example, could the neutrino be a Majorana particle which would mean that the mass term is generated by a higher dimensional operator with a large cutoff scale \cite{Weinberg:1979sa}. A very prominent mechanism that realizes this by introducing a heavy, right-handed, Majorana particle is the See-Saw mechanism \cite{Minkowski:1977sc, Gell-Mann:1979vob, Yanagida:1980xy, Mohapatra:1979ia, Mohapatra:2004zh}. Because a high cut-off scale is introduced such solutions to the neutrino mass problem are called UV solutions. 
\par 
Another solution to the neutrino mass problem offer theories in which the scale of gravity is lowered to TeV scale due to the presence of many additional light states, also called species \cite{Dvali:2007hz, Dvali:2007wp}. So far two models of this kind are known namely the Arkani-Hamed-Dimopoulos-Dvali (ADD) model \cite{Arkani-Hamed:1998jmv, Arkani-Hamed:1998sfv}, in which the light states are introduced by a Kaluza-Klein tower of gravitons, and the Dvali-Redi (DR) model with many copies of the SM \cite{Dvali:2009ne}. 
How small neutrino masses can be realized in the ADD and DR model was shown in \cite{Arkani-Hamed:1998wuz} and \cite{Dvali:2009ne}, respectively. Later it was demonstrated that ADD is a specific version of the many species approach \cite{Dvali:2007hz} and afterward that small Dirac masses for neutrinos is an inherent feature of this kind of theories \cite{Ettengruber:2022pxf} even though they were originally introduced to solve the hierarchy problem. Additionally, these theories have built-in solutions to the dark matter problem \cite{Arkani-Hamed:1999rvc, Dvali:2009ne, Dvali:2009fw}. Addressing this trinity of problems at the same time makes these theories candidates that are worth further investigation.
\par 
Let us stop here for a moment and see how these models fit into the bigger picture. The core motivation for these models comes from the following equation
\begin{equation}
    M_f \leq \frac{M_P}{\sqrt{N_{sp}}},
\end{equation}
where $M_P$ is the Planck mass, $M_f$ the fundamental scale of gravity and $N_{sp}$ is the number of additional light states. This equation gives us a way to solve the hierarchy problem of particle physics \cite{Dvali:2007hz, Dvali:2007wp} because if $N_{sp}$ is a large number of order $\mathcal{O}(10^{32})$ the fundamental scale of gravity can be lowered down to $TeV$ scale. This new scale $M_f$ in the context of string theory is often called the "species scale" \cite{Dvali:2008ec, Dvali:2009ks, Dvali:2010vm, Dvali:2012uq}. This scale, and not $M_P$ as one would naively think, marks the moment when gravity becomes non-perturbative and the description of Einstein's Gravity breaks down.

\par 
The fact that the separation of $M_f$ from $M_P$ influences the mass generation of the neutrino opens the exciting possibility to test the ADD and DR model with neutrino experiments that operate on energies far below the energy scale $M_f$ \cite{Dvali:1999cn, Dvali:2009ne, Ettengruber:2022pxf}. In this paper, we want to do exactly that. 

\par 

The ADD model was already the subject of previous research in \cite{Machado:2011jt,Machado:2011kt,Basto-Gonzalez:2012nel,Girardi:2014gna,Rodejohann:2014eka,Berryman:2016szd,Carena:2017qhd,Stenico:2018jpl,Arguelles:2019xgp,DUNE:2020fgq,Basto-Gonzalez:2021aus,Arguelles:2022xxa} and we will therefore focus on the DR model that has not been tested experimentally so far. We want to use the characteristic pattern for neutrino oscillations that the DR model predicts and search for it in neutrino experiments namely KamLAND \cite{KamLAND:2010fvi}, DayaBay \cite{DayaBay:2018yms, DayaBay:2016ssb}, MINOS/MINOS+ \cite{MINOS:2017cae}, NO$\nu$A \cite{NOvA:2018gge} and KATRIN \cite{KATRIN:2021uub}. We perform a global fit based on the publicly available data coming from these experiments and search for the imprints of the DR model in the data. Complementary theoretical considerations rooted in the cosmological history of the universe have been undertaken in \cite{Ettengruber:2023tac, Zander:2023jcu} and we compare our findings with these results.
In this work, we focus on terrestrial experiments to be independent of model-dependent cosmological scenarios. 
\par

This paper is organized as follows: In section \ref{Framework} we will introduce what the DR model actually is and how neutrino oscillations arise in this model. Then we carry on by describing our analysis strategy for the above-mentioned experiments in \ref{Analysis}. Finally, we present our results in \ref{Results} and give our conclusion and outlook in \ref{Conclusion}.

\section{Neutrino Oscillations with many additional light states}
\label{Framework}
 A particularly interesting case of theories with many additional species is the DR model where one introduces many SM copies that only interact gravitationally among each other \cite{Dvali:2009ne}. Such a scenario of many dark SM sectors could be easily realized in an extra-dimensional framework where the additional sectors are localized on displaced branes in the bulk as described in \cite{Arkani-Hamed:2016rle}. In such a situation the right-handed neutrino of every copy $\nu_R$ can play a special role because it can form the following Dirac term with their left-handed counterparts

    \begin{equation}
(HL)_i \lambda_{ij}\nu_{Rj} + h.c. .
\end{equation}
Here $H, L$ is the Higgs-/Lepton- Doublet $\lambda$ is a $N \times N$ Yukawa matrix, and the labels $i,j$ denote the different SM-copies. 

Because one introduces SM copies a full permutation symmetry among all neutrinos holds and this has the effect that the number of introduced parameters is very limited. These are: The number of neutrino copies $N$ and the Yukawa-matrix whose allowed structure is set to

\begin{equation}
    \lambda_{ij} =  \begin{pmatrix}
    a      & b      & \dots  & b \\
    b      & a      & \dots  & b \\
    \vdots & \vdots & \ddots & \vdots \\
    b      & b      & \dots  & a
    
    \end{pmatrix}  
    \label{Yukawa}.
\end{equation}
From unitarity, we get the following bound on $b$
\begin{equation}
    b \leq \frac{1}{\sqrt{N}},
\end{equation}
and we expect $a$ being of the same order because $a$ and $b$ are of the same nature. 
After diagonalizing the resulting mass matrix one gets the following expression for the neutrino made up of the mass eigenstates $ \nu_1^m $, $ \nu_H^m$
\begin{equation}
   \ket{ \nu_1} = \sqrt{\frac{N - 1}{N}} \ket{\nu_1^m} + \frac{1}{\sqrt{N}} \ket{\nu_H^m},
\end{equation}
with the corresponding eigenvalues
\begin{equation}
m_1 = (a-b)v,
\label{m_l}
\end{equation}
\begin{equation}
m_H = [a+(N-1)b]v,
\label{mH}
\end{equation}
where $v$ is the vacuum expectation value of the Higgs.
From (\ref{m_l}) we see that due to the suppression of the Yukawa-couplings with $\frac{1}{\sqrt{N}}$ the mass experiences the same suppression. This gives us a different view to explain the smallness of the neutrino mass compared to the Seesaw mechanism that offers a solution lying in the UV. Here the mass of the neutrino gets suppressed by the large number of additional light states. This gives us a solution to the neutrino mass problem that lies in the infrared (IR). This mechanism was originally described in \cite{Dvali:2009ne} and in \cite{Ettengruber:2022pxf} the setup was generalized to a three-flavor case leading to the expression for example for an electron neutrino
\begin{multline}
    \ket{\nu_{e}} = \sqrt{\frac{N-1}{N}} (U_{e1} \ket{\nu_1^m}  + U_{e2} \ket{\nu_2^m} + U_{e3} \ket{\nu_3^m} ) \\\\+  \frac{1}{\sqrt{N}} ( U_{e1}\ket{\nu_H^{m_1}}+ U_{e2} \ket{\nu_H^{m_2}} + U_{e3} \ket{\nu_H^{m_3}}),
    \label{manyspecies3flavor}
\end{multline}
and it was also shown, that by breaking the permutation symmetry among the copies by different vacuum expectation values, the masses of the heavy mass eigenstate (see \eqref{mH}) can be decoupled from $N$. The result is that the masses for the light and the heavy mass eigenstates are related via
\begin{equation}
    m_i = \mu m_H^i ,
\end{equation}
with $\mu$ being a factor that is the same for all three mass eigenstate pairs. This reduces the number of parameters beyond the SM down to just two parameters: $N$ and $\mu$. 

This is a specific feature of the neutrino extension of the DR model because the SM flavor mixing happens within one copy and the mixing among copies happens within the species (see \cite{Ettengruber:2022pxf}). Under this assumption, the above situation arises without any further input. This makes this theory highly predictive and offers us a smoking gun signature by relating the additional mass eigenstates to the SM ones by just one additional parameter. 

\begin{figure}
    \centering
    \includegraphics[width=\columnwidth]{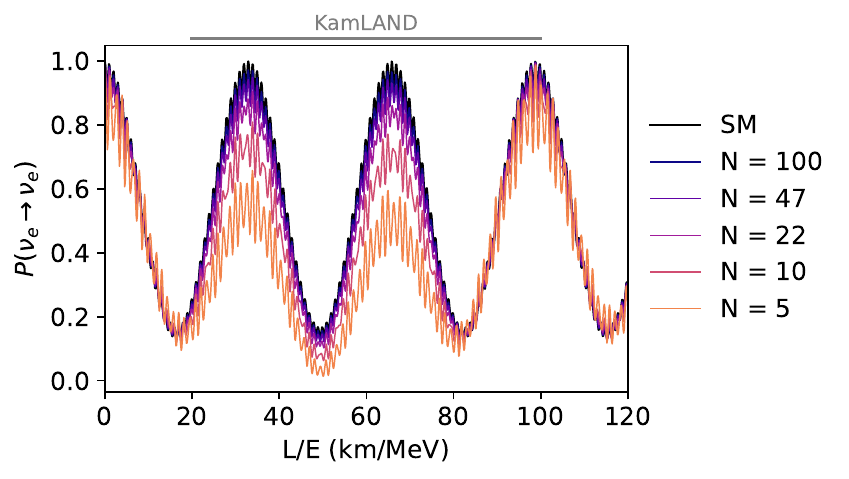}
    \caption{Example electron neutrino survival probabilities at an $L/E$ around the solar mass splitting for various numbers of extra species $N$ and massfactor $\mu = 5$.
}
    \label{fig:Kamland}
\end{figure}

\begin{figure}
    \centering
    \includegraphics[width=\columnwidth]{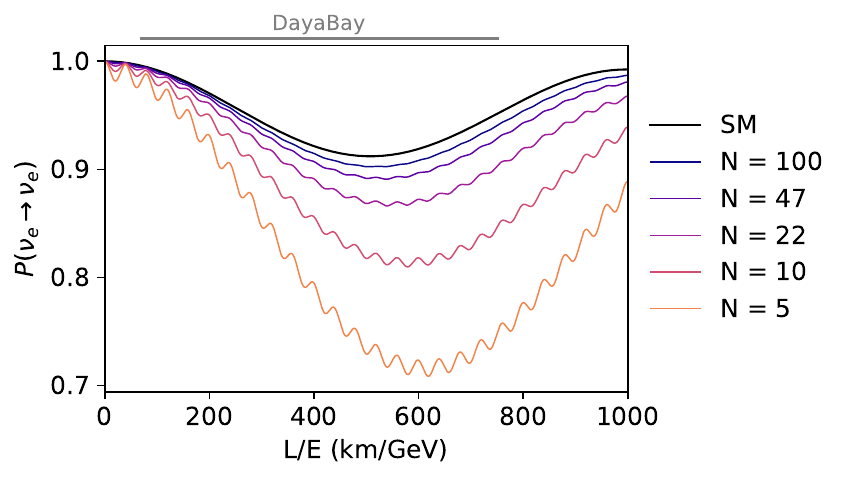}
    \caption{Example electron neutrino survival probabilities at an $L/E$ around the atmospheric mass splitting for various numbers of extra species $N$ and massfactor $\mu = 5$.}
    \label{fig:Daya}
\end{figure}

\begin{figure}
    \centering
    \includegraphics[width=\columnwidth]{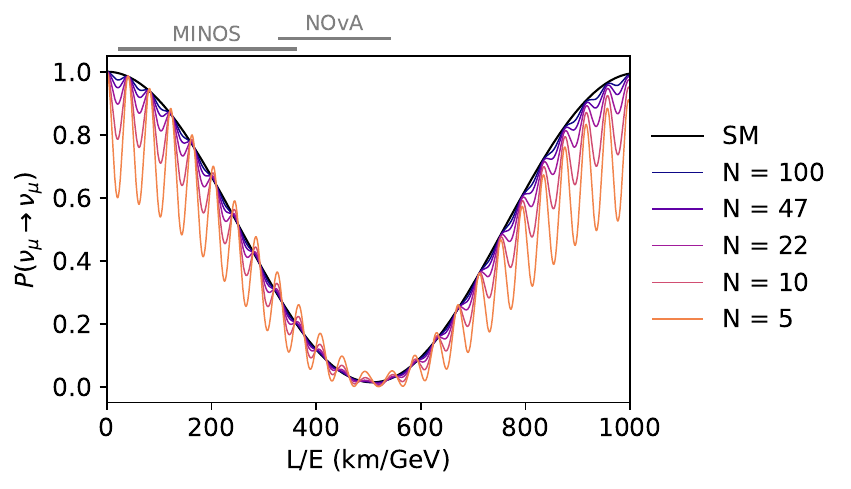}
    \caption{Example muon neutrino survival probabilities at an $L/E$ around the atmospheric mass splitting for various numbers of extra species $N$ and massfactor $\mu = 5$.}
    \label{fig:MinosNova}
\end{figure}

From the expression (\ref{manyspecies3flavor}) one can derive now the survival probability for such an electron neutrino with energy $E$ in the following way
\begin{multline}
     P(\nu_{e} \rightarrow \nu_{e}) = \left(\frac{N-1}{N}\right)^2 \sum_{i=1}^3 \sum_{j=1}^3 |U_{e i}|^2 |U_{e j}|^2 e^{\frac{i(m_i^2 - m_j^2)t}{2E}}\\\\ +  \frac{N-1}{N^2} \sum_{i=1}^3 \sum_{j=4}^6 |U_{e i}|^2 |U_{e j}|^2 e^{\frac{i(m_i^2 - m_j^2)t}{2E}} \\\\ +\frac{N-1}{N^2} \sum_{i=4}^6 \sum_{j=1}^3 |U_{e i}|^2 |U_{e j}|^2 e^{\frac{i(m_i^2 - m_j^2)t}{2E}} \\\\ + \frac{1}{N^2}\sum_{i=4}^6 \sum_{j=4}^6 |U_{e i}|^2 |U_{e j}|^2 e^{\frac{i(m_i^2 - m_j^2)t}{2E}}.
     \label{SurvProb}
\end{multline}
Example oscillation patterns of this formula for the choice of a few parameter values are depicted in the Figs.~\ref{fig:Kamland}, \ref{fig:Daya} and \ref{fig:MinosNova} in the $L/E$ regimes of interest.
\par
But not just the oscillator behavior of neutrinos get affected in this theory but the additional mass eigenstates will influence the effective mass of interacting neutrino states. This offers us another way to restrict the parameter space of the DR model by using the upper bound on the electron neutrino mass via the following formula
\begin{multline}
    m_{\nu_e}^2 =  m_{\rm lightest}^2 \left( \frac{N+\mu^2-1}{N}\right) + \frac{N-1}{N} ( \Delta m_{12}^2 U_{e2}^2 + \\\\   \Delta m_{13}^2 U_{e3}^2 ) + \frac{\mu^2}{N} \left( \Delta m_{12}^2 U_{e2}^2+  \Delta m_{13}^2 U_{e3}^2 \right).
    \label{neutrinomass}
\end{multline}
So we have two types of experiments to determine our DR parameters, neutrino oscillation experiments, and neutrino mass detection experiments. In principle, one could also use data from neutrino mass sum probes coming from cosmology. But as we said, in this paper we want to restrict ourselves to experiments that are independent from the cosmological model.
\par 
The predictive power of the theory discussed above becomes evident when one compares it with a general ansatz where one sets the number of additional sterile neutrinos $n_s = 3$. In such an approach the number of independent physical mixing angles is $3(n_s + 1) = 12$ and Dirac phases are $2n_s+ 1 = 7$. On top we have three additional masses for the sterile neutrinos, $m^s_1$, $m^s_2$, $m^s_3$ \cite{Barry:2011wb}. The total number of independent parameters of a 3+3 approach is 22. The DR model has two BSM parameters, $N$, and $\mu$. Because $\mu$ relates the masses of the SM neutrinos with the sterile neutrinos, another parameter of interest is the mass of the lightest neutrino $m_{\textrm{lightest}}$. This gives us a number of 3 parameters that are of additional interest for neutrino oscillations on top of the usual SM parameters. The total number of parameters to fit neutrino oscillations is then 9. Obviously, the DR model offers a more minimal framework to perform a BSM neutrino fit than a general 3+3 neutrino fit.

\section{Data Analysis}
\label{Analysis}

Overall, our analysis has the following free parameters:
\begin{equation}
  \{\delta_{CP}, \theta_{12}, \theta_{13}, \theta_{23}, \Delta m_{12}^2, \Delta m_{13}^2, m_{\textrm{lightest}}, N, \mu \},
   \label{parameters}
\end{equation}
where $\delta_{CP}, \theta_{12}, \theta_{13}, \theta_{23}$ are the well known parameters of the PMNS matrix and $\Delta m_{12}^2, \Delta m_{13}^2$ the differences of the mass eigenstates.

To determine the values of these parameters we perform a combined maximum likelihood fit. Because we analyze neutrino data for the first time in the DR regime we do not have prior knowledge of the parameters shown in equation (\ref{parameters}). Therefore, we have to perform an independent analysis of the available experimental data and cannot rely on previous SM global fits. Also, because every SM mixing angle experiences a correction by this model and the additional mass splittings can in principle take very different values our statistical analysis has to perform a fit for all 9 free parameters. 

The answer to the question of which experiment to include in our analysis we based on the fact that the neutrino oscillation experiments measure different combinations of the fraction $\frac{L}{E}$, with $L$ being the base lengths of the experiment and $E$ the energies of the neutrinos. Including a variety of different experiments has two effects. First, we can choose for every SM parameter at least one experiment that restrains this parameter. Secondly, a broad range of the different experiments allows us also to search for deviations in SM oscillations coming from very small to quite large mass splittings between the SM mass eigenstates and the BSM ones.

\par 
Now let us discuss our specific choice of experiments and what we expect how the different experiments contribute to the resulting exclusion limits of the BSM parameters.

 Maybe the intuitively easiest experiment to understand is KATRIN. The additional mass eigenstates will contribute to the effective mass of the electron neutrino and the model parameters controlling this mass are $m_{\textrm{lightest}}$ and $\mu$. Meanwhile, oscillations experiments are also sensitive to $\mu$, KATRIN is the only experiment in our analysis that can restrain $m_{\textrm{lightest}}$. This is important because the masses of the SM states are uniquely determined after $m_{\textrm{lightest}}$ is fixed and $\mu$ is then relating the SM masses to the BSM ones. Therefore, for large $\mu$ the BSM masses are higher and the electron neutrino mass gets stronger influenced. Due to this, we expect KATRIN to become most relevant in the regime where $\mu$ has large values. 

\par 
The KamLAND experiment is necessary for our analysis because it restricts the SM parameters $\Delta m_{12}^2$ and $\theta_{12}$. The expected sensitivity for BSM parameters we can estimate if we look at Fig. \ref{fig:Kamland} where we see that the deviations from the SM oscillations become significant for $N \leq 10$. 
\par 
DayaBay on the other hand is much more sensitive to $N$ as we see from Fig. \ref{fig:Daya}. Together with the fact that this experiment provides us with very high statistics in its data, we expect that DayaBay will contribute significantly to our final exclusion limit for the BSM parameters. Additionally it is restricting $\theta_{13}$ and $\Delta m_{13}^2$. 
\par 
MINOS and NO$\nu$A are complementary. Because No$\nu$a is located exactly at the oscillation maximum it provides very good sensitivity for SM parameters $\theta_{23}$ and $\Delta m_{13}^2$ but at the same time it is lacking sensitivity to the BSM parameters (see Fig. \ref{fig:MinosNova}). Minos on the other hand is located off maximum which is suited for searching for BSM contributions to the oscillation pattern. 
\par

Putting the pieces together we can define the following likelihood
\begin{multline}
    \mathcal{L}_\mathrm{comb} = \mathcal{L}_\mathrm{KATRIN} \times \mathcal{L}_\mathrm{MINOS}\times \mathcal{L}_\mathrm{KamLAND} \\\\ \times \mathcal{L}_\mathrm{DayaBay}\times \mathcal{L}_\mathrm{NO\nu A},
\end{multline}
where we treated every data set independently which allows us to construct $\mathcal{L}_{\textrm{comb}}$ as a product of all the single likelihoods of the experiments. For our statistical analysis, we use a likelihood ratio test statistic. In case the alternate hypothesis (DR model) would be preferred over the null hypothesis (SM) with a significance greater than $3\sigma$, we would investigate a signal. Otherwise, we would set exclusion limits on the parameter $N$ as a function of $\mu$, while profiling over all remaining parameters.

\par 

In this analysis, public data was used that was given in the papers \cite{KamLAND:2010fvi,DayaBay:2018yms, DayaBay:2016ssb, MINOS:2017cae, NOvA:2018gge, KATRIN:2021uub}. For each experiment, a fit of the SM model was performed and compared with the findings of the articles to ensure the validity of the analysis using the DR model. A deeper summary of how each experiment was analyzed is given in the Appendix.

\section{Results}
\label{Results}

The best fitting DR hypothesis assuming normal ordering (NO) yields a value in log-likelihood units of 4.37 better than the standard model fit, while for the inverted ordering (IO) the difference equates to 2.49 log-likelihood units. These numbers, assuming Wilk's theorem and considering the three additional degrees of freedom ($N, \mu, m_{\textrm{lightest}}$), correspond to a significance of $1.8\sigma$ and $0.97\sigma$, respectively, for the NO and IO case.
Since neither of these are at a significance $>3\sigma$, we proceed to set exclusion limits.

\begin{figure*}
    \centering
    \includegraphics[width=0.8\textwidth]{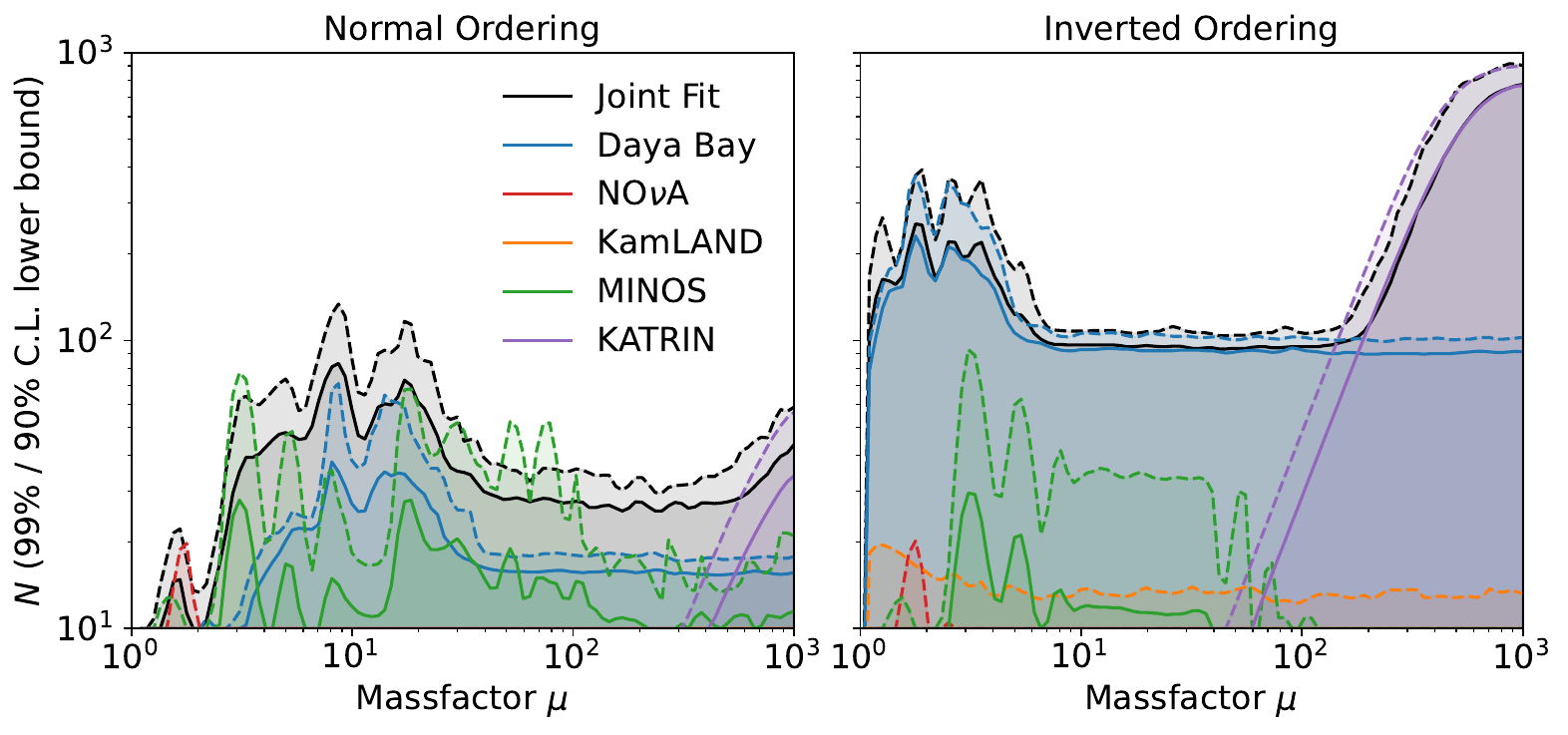}
    \caption{Lower bounds on the number of species $N$ as a function of the massfactor $\mu$ for the normal and inverted neutrino mass ordering, respectively. The solid and dashed lines denote the 90\% and 99\% asymptotic confidence levels (CL), the shaded areas are excluded. The colorful lines represent fits of individual datasets, and the black lines result from a combined fit of all four datasets.}
    \label{IOSpecies}
\end{figure*}

The resulting limits are shown in Figure \ref{IOSpecies} for NO and IO, respectively.
First, we see that in the IO case, we can give a lower bound on $N > \mathcal{O}(10^2)$ quite consistently over the range of $\mu$. The exclusion limit for $\mu< 10^2$ is set by the oscillations experiments, meanwhile for $\mu > 10^2$ the exclusion is basically dominated by KATRIN as expected. Because oscillations experiments measure the survival probability \eqref{SurvProb} these experiments test a beyond SM contribution that scales in the first order as $\frac{1}{N}$. This means that for $N > \mathcal{O}(10^2)$ the corrections coming from the DR model are of order $<1\%$ which is consistent with reported experimental measurements that have uncertainties around $1-5 \%$. The situation is more complex in the NO case. We see in general that the bounds on $N$ are weaker than in the IO case. The significance of KATRIN is reduced compared to the IO case and starts affecting the exclusion limit for $\mu >300$. Nevertheless, for higher values of $\mu$ it shows a similar behavior as in the NO case but with a reduced sensitivity. In the parameterspace where $\mu > 2$ one can give a rough bound on $N > 30$. If $\mu \leq 2$ then one can see that experiments start losing their sensitivity to this model because the mass-splitting for the BSM states becomes too small to be resolved by experiments. 
\par 
If we compare our numerical results with what we would have expected by the oscillograms as described in the previous section we see that our expectations agree with our results.

\section{Conclusion}
\label{Conclusion}
In this work, we presented the first experimental test of additional neutrino copies using experimental data. We concluded that depending on the mass hierarchy realized in nature we can set a lower limit on the number of neutrino copies as $N > \mathcal{O}(30)$ for NO and $N > \mathcal{O}(100)$ for IO over a wide range of the parameter space. 
\par 
Using the fact that the Dirac operator allows communication via the right-handed neutrino between the sectors one can also add cosmological considerations to give lower bounds on the number of neutrino species \cite{Zander:2023jcu}. Compared to these bounds the bounds resulting from experiments are weaker but more robust due to the fact that our results do not depend on the cosmological history beyond BBN.
\par 

Therefore our results show, that neutrino experiments are especially suited for testing models with additional neutrino species because compared to other physics we can give a lower bound on $N$ meanwhile LHC \cite{Dvali:2007hz, Dvali:2007wp} and axion physics \cite{Ettengruber:2023tac} give an upper bound on the number of species. 

The complementary nature of neutrino experiments operating in the IR to UV experiments is quite exciting and future experiments like JUNO \cite{JUNO:2015sjr} and DUNE \cite{DUNE:2020lwj} are designed to improve our knowledge about the lepton mixing parameters by one order of magnitude which will give us the possibility to close the open window of the parameter space even further. 

\par 
Together with theoretical considerations \cite{Ettengruber:2023tac, Zander:2023jcu}, UV and IR experiments can help us to restrict the possible range of $N_{sp}$ and reveal where we could expect the true scale of Quantum Gravity. 

\section*{Acknowledgments}

We are thankful for our discussions with Allen Caldwell and Gia Dvali. This work has been supported by the Deutsche
Forschungsgemeinschaft (DFG, German Research Foundation) under the Sonderforschungsbereich (Collaborative Research Center) SFB1258 ‘Neutrinos and Dark Matter in Astro- and Particle Physics’.

\section{Appendix}
In the appendix, we describe in more detail how the different experiments have been analyzed. 
\subsection{DayaBay}
DayaBay was a reactor neutrino experiment located in China to measure the parameters $\theta_{13}$ and $\Delta m_{32}^2$. Together with \cite{DayaBay:2018yms} the collaboration released their data set of 26 data points which we analyzed in this work. Because DayaBay's neutrino flux is sourced by several nuclear reactors which have different baselengths to the experimental halls the detectors are placed in, we calculated the contribution of every reactor to the overall flux at the experimental site. DayaBay consists of experimental halls close to the reactor to determine the predicted flux in the far experimental hall. The data of the far experimental hall (EH3) was used in this analysis. The geometric averaged baseline to EH3 is around 1663 m.
In order to incorporate the systematical uncertainties of the experiment we took the covariance matrix published in \cite{DayaBay:2016ssb} and scaled the general covariance matrix accordingly to the data which is analyzed here. 
Performing a fit for standard mixing parameters, we find that our values are well consistent with those reported by the collaboration. 
\subsection{KamLAND}
KamLAND was a reactor experiment based in Japan designed to measure $\theta_{12}$ and $\Delta m_{12}^2$.
We averaged the survival probability of the neutrino flux over all baselines of the reactors placed in Japan while neglecting the contribution from South Korea which is around $5\%$ and the world contribution which is around $1\%$. Our analysis is based on the publication \cite{KamLAND:2010fvi} that also includes a non-oscillated spectrum of the neutrino flux at the experiment. The necessary information was extracted from Fig.~1 in this publication. Due to lacking public information, we could not include a full covariance matrix and resorted to using a diagonal covariance matrix which we constructed by using the uncertainties of the measured events per bin.
In our final analysis, we incorporated 17 data points. The energy resolution was approximated by the bin width. Our fit results are well consistent with those reported by the collaboration.

\subsection{MINOS}
MINOS was an accelerator muon neutrino experiment located at Fermilab in the U.S. that operates slightly off maximum of the atmospheric mass splitting regime (see Fig. \ref{fig:MinosNova}). In principle, it can be used to determine $\theta_{23}$ and $\Delta m_{13}^2$ even though due to the energy range it is not optimal, but provides excellent data for searching for BSM signals in the oscillation pattern. This makes this experiment of particular interest for our analysis. We use the far detector (FD) CC and NC data from MINOS and MINOS+. The simulation templates, smearing matrices, covariance matrices, as well as the observed counts are provided in the data release accompanying the publication \cite{MINOS:2017cae}. We implement our analysis by replicating the provided reference implementation. Our fit results are well consistent with those reported by the collaboration. 

\subsection{NO$\nu$A}
Also, NO$\nu$A uses the same muon neutrino beam as MINOS but this experiment is located off-axis, resulting in maximum mixing if the atmospheric mass splitting regime (see Fig. \ref{fig:MinosNova}). Therefore, it is optimized to restrict the SM parameters $\theta_{23}$ and $\Delta m_{13}^2$.
We use the muon disappearance data from \cite{NOvA:2021nfi}. Forward and reversed horn current data is included in our analysis. A detector response is implemented via a smearing function defined according to the resolution specified in Table 2 in  \cite{NOvA:2021nfi}. Our fit results are well consistent with those reported by the collaboration.

\subsection{KATRIN}

Another relevant type of experiment is the direct measurement of the neutrino mass. The leading experiment for this is KATRIN which analyzes the spectrum of beta decays. In \cite{KATRIN:2021uub} the collaboration also performed a Bayesian analysis and reported a posterior on the mass of the electron neutrino. Because a flat prior on the $m_{\rm lightest}$ was used to calculate this result, we can easily interpret this posterior as our likelihood. We approximated the posterior distribution with a truncated normal distribution and calculated the predicted neutrino mass of the DR model with \eqref{neutrinomass}. In this analysis, we just evaluated the likelihood as described above and did not take into account the change in the shape of the energy distribution which would be caused by additional mass states in the expression for the flavor states. This is usually done in sterile neutrino searches with KATRIN and could still be improved in our analysis.  

\setlength{\bibsep}{5pt}

\bibliographystyle{utphys}
\bibliography{refs}

\providecommand{\href}[2]{#2}\begingroup\raggedright\begin{thebibliography}{10}

\bibitem{Weinberg:1979sa}
S.~Weinberg, ``{Baryon and Lepton Nonconserving Processes},''
  \href{http://dx.doi.org/10.1103/PhysRevLett.43.1566}{{\em Phys. Rev. Lett.}
  {\bfseries 43} (1979) 1566--1570}.

\bibitem{Minkowski:1977sc}
P.~Minkowski, ``{$\mu \to e\gamma$ at a Rate of One Out of $10^{9}$ Muon
  Decays?},'' \href{http://dx.doi.org/10.1016/0370-2693(77)90435-X}{{\em Phys.
  Lett. B} {\bfseries 67} (1977) 421--428}.

\bibitem{Gell-Mann:1979vob}
M.~Gell-Mann, P.~Ramond, and R.~Slansky, ``{Complex Spinors and Unified
  Theories},'' {\em Conf. Proc. C} {\bfseries 790927} (1979) 315--321,
  \href{http://arxiv.org/abs/1306.4669}{{\ttfamily arXiv:1306.4669 [hep-th]}}.

\bibitem{Yanagida:1980xy}
T.~Yanagida, ``{Horizontal Symmetry and Masses of Neutrinos},''
  \href{http://dx.doi.org/10.1143/PTP.64.1103}{{\em Prog. Theor. Phys.}
  {\bfseries 64} (1980) 1103}.

\bibitem{Mohapatra:1979ia}
R.~N. Mohapatra and G.~Senjanovic, ``{Neutrino Mass and Spontaneous Parity
  Nonconservation},'' \href{http://dx.doi.org/10.1103/PhysRevLett.44.912}{{\em
  Phys. Rev. Lett.} {\bfseries 44} (1980) 912}.

\bibitem{Mohapatra:2004zh}
R.~N. Mohapatra, \href{http://dx.doi.org/10.1142/9789812702210_0003}{``{Seesaw
  mechanism and its implications},''} in {\em {SEESAW25: International
  Conference on the Seesaw Mechanism and the Neutrino Mass}}, pp.~29--44.
\newblock 12, 2004.
\newblock \href{http://arxiv.org/abs/hep-ph/0412379}{{\ttfamily
  arXiv:hep-ph/0412379}}.

\bibitem{Dvali:2007hz}
G.~Dvali, ``{Black Holes and Large N Species Solution to the Hierarchy
  Problem},'' \href{http://dx.doi.org/10.1002/prop.201000009}{{\em Fortsch.
  Phys.} {\bfseries 58} (2010) 528--536},
  \href{http://arxiv.org/abs/0706.2050}{{\ttfamily arXiv:0706.2050 [hep-th]}}.

\bibitem{Dvali:2007wp}
G.~Dvali and M.~Redi, ``{Black Hole Bound on the Number of Species and Quantum
  Gravity at LHC},'' \href{http://dx.doi.org/10.1103/PhysRevD.77.045027}{{\em
  Phys. Rev. D} {\bfseries 77} (2008) 045027},
  \href{http://arxiv.org/abs/0710.4344}{{\ttfamily arXiv:0710.4344 [hep-th]}}.

\bibitem{Arkani-Hamed:1998jmv}
N.~Arkani-Hamed, S.~Dimopoulos, and G.~R. Dvali, ``{The Hierarchy problem and
  new dimensions at a millimeter},''
  \href{http://dx.doi.org/10.1016/S0370-2693(98)00466-3}{{\em Phys. Lett. B}
  {\bfseries 429} (1998) 263--272},
  \href{http://arxiv.org/abs/hep-ph/9803315}{{\ttfamily arXiv:hep-ph/9803315}}.

\bibitem{Arkani-Hamed:1998sfv}
N.~Arkani-Hamed, S.~Dimopoulos, and G.~R. Dvali, ``{Phenomenology, astrophysics
  and cosmology of theories with submillimeter dimensions and TeV scale quantum
  gravity},'' \href{http://dx.doi.org/10.1103/PhysRevD.59.086004}{{\em Phys.
  Rev. D} {\bfseries 59} (1999) 086004},
  \href{http://arxiv.org/abs/hep-ph/9807344}{{\ttfamily arXiv:hep-ph/9807344}}.

\bibitem{Dvali:2009ne}
G.~Dvali and M.~Redi, ``{Phenomenology of $10^{32}$ Dark Sectors},''
  \href{http://dx.doi.org/10.1103/PhysRevD.80.055001}{{\em Phys. Rev. D}
  {\bfseries 80} (2009) 055001},
  \href{http://arxiv.org/abs/0905.1709}{{\ttfamily arXiv:0905.1709 [hep-ph]}}.

\bibitem{Arkani-Hamed:1998wuz}
N.~Arkani-Hamed, S.~Dimopoulos, G.~R. Dvali, and J.~March-Russell, ``{Neutrino
  masses from large extra dimensions},''
  \href{http://dx.doi.org/10.1103/PhysRevD.65.024032}{{\em Phys. Rev. D}
  {\bfseries 65} (2001) 024032},
  \href{http://arxiv.org/abs/hep-ph/9811448}{{\ttfamily arXiv:hep-ph/9811448}}.

\bibitem{Ettengruber:2022pxf}
M.~Ettengruber, ``{Neutrino physics in TeV scale gravity theories},''
  \href{http://dx.doi.org/10.1103/PhysRevD.106.055028}{{\em Phys. Rev. D}
  {\bfseries 106} no.~5, (2022) 055028},
  \href{http://arxiv.org/abs/2206.00034}{{\ttfamily arXiv:2206.00034
  [hep-ph]}}.

\bibitem{Arkani-Hamed:1999rvc}
N.~Arkani-Hamed, S.~Dimopoulos, G.~R. Dvali, and N.~Kaloper, ``{Many fold
  universe},'' \href{http://dx.doi.org/10.1088/1126-6708/2000/12/010}{{\em
  JHEP} {\bfseries 12} (2000) 010},
  \href{http://arxiv.org/abs/hep-ph/9911386}{{\ttfamily arXiv:hep-ph/9911386}}.

\bibitem{Dvali:2009fw}
G.~Dvali, I.~Sawicki, and A.~Vikman, ``{Dark Matter via Many Copies of the
  Standard Model},''
  \href{http://dx.doi.org/10.1088/1475-7516/2009/08/009}{{\em JCAP} {\bfseries
  08} (2009) 009}, \href{http://arxiv.org/abs/0903.0660}{{\ttfamily
  arXiv:0903.0660 [hep-th]}}.

\bibitem{Dvali:2008ec}
G.~Dvali and C.~Gomez, ``{Quantum Information and Gravity Cutoff in Theories
  with Species},'' \href{http://dx.doi.org/10.1016/j.physletb.2009.03.024}{{\em
  Phys. Lett. B} {\bfseries 674} (2009) 303--307},
  \href{http://arxiv.org/abs/0812.1940}{{\ttfamily arXiv:0812.1940 [hep-th]}}.

\bibitem{Dvali:2009ks}
G.~Dvali and D.~Lust, ``{Evaporation of Microscopic Black Holes in String
  Theory and the Bound on Species},''
  \href{http://dx.doi.org/10.1002/prop.201000008}{{\em Fortsch. Phys.}
  {\bfseries 58} (2010) 505--527},
  \href{http://arxiv.org/abs/0912.3167}{{\ttfamily arXiv:0912.3167 [hep-th]}}.

\bibitem{Dvali:2010vm}
G.~Dvali and C.~Gomez, ``{Species and Strings},''
  \href{http://arxiv.org/abs/1004.3744}{{\ttfamily arXiv:1004.3744 [hep-th]}}.

\bibitem{Dvali:2012uq}
G.~Dvali, C.~Gomez, and D.~Lust, ``{Black Hole Quantum Mechanics in the
  Presence of Species},'' \href{http://dx.doi.org/10.1002/prop.201300002}{{\em
  Fortsch. Phys.} {\bfseries 61} (2013) 768--778},
  \href{http://arxiv.org/abs/1206.2365}{{\ttfamily arXiv:1206.2365 [hep-th]}}.

\bibitem{Dvali:1999cn}
G.~R. Dvali and A.~Y. Smirnov, ``{Probing large extra dimensions with
  neutrinos},'' \href{http://dx.doi.org/10.1016/S0550-3213(99)00574-X}{{\em
  Nucl. Phys. B} {\bfseries 563} (1999) 63--81},
  \href{http://arxiv.org/abs/hep-ph/9904211}{{\ttfamily arXiv:hep-ph/9904211}}.

\bibitem{Machado:2011jt}
P.~A.~N. Machado, H.~Nunokawa, and R.~Zukanovich~Funchal, ``{Testing for Large
  Extra Dimensions with Neutrino Oscillations},''
  \href{http://dx.doi.org/10.1103/PhysRevD.84.013003}{{\em Phys. Rev. D}
  {\bfseries 84} (2011) 013003},
  \href{http://arxiv.org/abs/1101.0003}{{\ttfamily arXiv:1101.0003 [hep-ph]}}.

\bibitem{Machado:2011kt}
P.~A.~N. Machado, H.~Nunokawa, F.~A.~P. dos Santos, and R.~Z. Funchal, ``{Bulk
  Neutrinos as an Alternative Cause of the Gallium and Reactor Anti-neutrino
  Anomalies},'' \href{http://dx.doi.org/10.1103/PhysRevD.85.073012}{{\em Phys.
  Rev. D} {\bfseries 85} (2012) 073012},
  \href{http://arxiv.org/abs/1107.2400}{{\ttfamily arXiv:1107.2400 [hep-ph]}}.

\bibitem{Basto-Gonzalez:2012nel}
V.~S. Basto-Gonzalez, A.~Esmaili, and O.~L.~G. Peres, ``{Kinematical Test of
  Large Extra Dimension in Beta Decay Experiments},''
  \href{http://dx.doi.org/10.1016/j.physletb.2012.11.048}{{\em Phys. Lett. B}
  {\bfseries 718} (2013) 1020--1023},
  \href{http://arxiv.org/abs/1205.6212}{{\ttfamily arXiv:1205.6212 [hep-ph]}}.

\bibitem{Girardi:2014gna}
I.~Girardi and D.~Meloni, ``{Constraining new physics scenarios in neutrino
  oscillations from Daya Bay data},''
  \href{http://dx.doi.org/10.1103/PhysRevD.90.073011}{{\em Phys. Rev. D}
  {\bfseries 90} no.~7, (2014) 073011},
  \href{http://arxiv.org/abs/1403.5507}{{\ttfamily arXiv:1403.5507 [hep-ph]}}.

\bibitem{Rodejohann:2014eka}
W.~Rodejohann and H.~Zhang, ``{Signatures of Extra Dimensional Sterile
  Neutrinos},'' \href{http://dx.doi.org/10.1016/j.physletb.2014.08.035}{{\em
  Phys. Lett. B} {\bfseries 737} (2014) 81--89},
  \href{http://arxiv.org/abs/1407.2739}{{\ttfamily arXiv:1407.2739 [hep-ph]}}.

\bibitem{Berryman:2016szd}
J.~M. Berryman, A.~de~Gouv\^ea, K.~J. Kelly, O.~L.~G. Peres, and Z.~Tabrizi,
  ``{Large, Extra Dimensions at the Deep Underground Neutrino Experiment},''
  \href{http://dx.doi.org/10.1103/PhysRevD.94.033006}{{\em Phys. Rev. D}
  {\bfseries 94} no.~3, (2016) 033006},
  \href{http://arxiv.org/abs/1603.00018}{{\ttfamily arXiv:1603.00018
  [hep-ph]}}.

\bibitem{Carena:2017qhd}
M.~Carena, Y.-Y. Li, C.~S. Machado, P.~A.~N. Machado, and C.~E.~M. Wagner,
  ``{Neutrinos in Large Extra Dimensions and Short-Baseline $\nu_e$
  Appearance},'' \href{http://dx.doi.org/10.1103/PhysRevD.96.095014}{{\em Phys.
  Rev. D} {\bfseries 96} no.~9, (2017) 095014},
  \href{http://arxiv.org/abs/1708.09548}{{\ttfamily arXiv:1708.09548
  [hep-ph]}}.

\bibitem{Stenico:2018jpl}
G.~V. Stenico, D.~V. Forero, and O.~L.~G. Peres, ``{A Short Travel for
  Neutrinos in Large Extra Dimensions},''
  \href{http://dx.doi.org/10.1007/JHEP11(2018)155}{{\em JHEP} {\bfseries 11}
  (2018) 155}, \href{http://arxiv.org/abs/1808.05450}{{\ttfamily
  arXiv:1808.05450 [hep-ph]}}.

\bibitem{Arguelles:2019xgp}
C.~A. Arg\"uelles {\em et~al.}, ``{New opportunities at the next-generation
  neutrino experiments I: BSM neutrino physics and dark matter},''
  \href{http://dx.doi.org/10.1088/1361-6633/ab9d12}{{\em Rept. Prog. Phys.}
  {\bfseries 83} no.~12, (2020) 124201},
  \href{http://arxiv.org/abs/1907.08311}{{\ttfamily arXiv:1907.08311
  [hep-ph]}}.

\bibitem{DUNE:2020fgq}
{\bfseries DUNE} Collaboration, B.~Abi {\em et~al.}, ``{Prospects for beyond
  the Standard Model physics searches at the Deep Underground Neutrino
  Experiment},'' \href{http://dx.doi.org/10.1140/epjc/s10052-021-09007-w}{{\em
  Eur. Phys. J. C} {\bfseries 81} no.~4, (2021) 322},
  \href{http://arxiv.org/abs/2008.12769}{{\ttfamily arXiv:2008.12769
  [hep-ex]}}.

\bibitem{Basto-Gonzalez:2021aus}
V.~S. Basto-Gonzalez, D.~V. Forero, C.~Giunti, A.~A. Quiroga, and C.~A. Ternes,
  ``{Short-baseline oscillation scenarios at JUNO and TAO},''
  \href{http://dx.doi.org/10.1103/PhysRevD.105.075023}{{\em Phys. Rev. D}
  {\bfseries 105} no.~7, (2022) 075023},
  \href{http://arxiv.org/abs/2112.00379}{{\ttfamily arXiv:2112.00379
  [hep-ph]}}.

\bibitem{Arguelles:2022xxa}
C.~A. Arg\"uelles {\em et~al.}, ``{Snowmass White Paper: Beyond the Standard
  Model effects on Neutrino Flavor},'' in {\em {2022 Snowmass Summer Study}}.
\newblock 3, 2022.
\newblock \href{http://arxiv.org/abs/2203.10811}{{\ttfamily arXiv:2203.10811
  [hep-ph]}}.

\bibitem{KamLAND:2010fvi}
{\bfseries KamLAND} Collaboration, A.~Gando {\em et~al.}, ``{Constraints on
  $\theta_{13}$ from A Three-Flavor Oscillation Analysis of Reactor
  Antineutrinos at KamLAND},''
  \href{http://dx.doi.org/10.1103/PhysRevD.83.052002}{{\em Phys. Rev. D}
  {\bfseries 83} (2011) 052002},
  \href{http://arxiv.org/abs/1009.4771}{{\ttfamily arXiv:1009.4771 [hep-ex]}}.

\bibitem{DayaBay:2018yms}
{\bfseries Daya Bay} Collaboration, D.~Adey {\em et~al.}, ``{Measurement of the
  Electron Antineutrino Oscillation with 1958 Days of Operation at Daya Bay},''
  \href{http://dx.doi.org/10.1103/PhysRevLett.121.241805}{{\em Phys. Rev.
  Lett.} {\bfseries 121} no.~24, (2018) 241805},
  \href{http://arxiv.org/abs/1809.02261}{{\ttfamily arXiv:1809.02261
  [hep-ex]}}.

\bibitem{DayaBay:2016ssb}
{\bfseries Daya Bay} Collaboration, F.~P. An {\em et~al.}, ``{Improved
  Measurement of the Reactor Antineutrino Flux and Spectrum at Daya Bay},''
  \href{http://dx.doi.org/10.1088/1674-1137/41/1/013002}{{\em Chin. Phys. C}
  {\bfseries 41} no.~1, (2017) 013002},
  \href{http://arxiv.org/abs/1607.05378}{{\ttfamily arXiv:1607.05378
  [hep-ex]}}.

\bibitem{MINOS:2017cae}
{\bfseries MINOS+} Collaboration, P.~Adamson {\em et~al.}, ``{Search for
  sterile neutrinos in MINOS and MINOS+ using a two-detector fit},''
  \href{http://dx.doi.org/10.1103/PhysRevLett.122.091803}{{\em Phys. Rev.
  Lett.} {\bfseries 122} no.~9, (2019) 091803},
  \href{http://arxiv.org/abs/1710.06488}{{\ttfamily arXiv:1710.06488
  [hep-ex]}}.

\bibitem{NOvA:2018gge}
{\bfseries NOvA} Collaboration, M.~A. Acero {\em et~al.}, ``{New constraints on
  oscillation parameters from $\nu_e$ appearance and $\nu_\mu$ disappearance in
  the NOvA experiment},''
  \href{http://dx.doi.org/10.1103/PhysRevD.98.032012}{{\em Phys. Rev. D}
  {\bfseries 98} (2018) 032012},
  \href{http://arxiv.org/abs/1806.00096}{{\ttfamily arXiv:1806.00096
  [hep-ex]}}.

\bibitem{KATRIN:2021uub}
{\bfseries KATRIN} Collaboration, M.~Aker {\em et~al.}, ``{Direct neutrino-mass
  measurement with sub-electronvolt sensitivity},''
  \href{http://dx.doi.org/10.1038/s41567-021-01463-1}{{\em Nature Phys.}
  {\bfseries 18} no.~2, (2022) 160--166},
  \href{http://arxiv.org/abs/2105.08533}{{\ttfamily arXiv:2105.08533
  [hep-ex]}}.

\bibitem{Ettengruber:2023tac}
M.~Ettengruber and E.~Koutsangelas, ``{Consequences of Multiple Axions in
  Theories with Dark Yang-Mills Groups},''
  \href{http://arxiv.org/abs/2307.10298}{{\ttfamily arXiv:2307.10298
  [hep-ph]}}.

\bibitem{Zander:2023jcu}
A.~Zander, M.~Ettengruber, and P.~Eller, ``{How Many Dark Neutrino Sectors Does
  Cosmology Allow?},'' \href{http://arxiv.org/abs/2308.00798}{{\ttfamily
  arXiv:2308.00798 [hep-ph]}}.

\bibitem{Arkani-Hamed:2016rle}
N.~Arkani-Hamed, T.~Cohen, R.~T. D'Agnolo, A.~Hook, H.~D. Kim, and D.~Pinner,
  ``{Solving the Hierarchy Problem at Reheating with a Large Number of Degrees
  of Freedom},'' \href{http://dx.doi.org/10.1103/PhysRevLett.117.251801}{{\em
  Phys. Rev. Lett.} {\bfseries 117} no.~25, (2016) 251801},
  \href{http://arxiv.org/abs/1607.06821}{{\ttfamily arXiv:1607.06821
  [hep-ph]}}.

\bibitem{Barry:2011wb}
J.~Barry, W.~Rodejohann, and H.~Zhang, ``{Light Sterile Neutrinos: Models and
  Phenomenology},'' \href{http://dx.doi.org/10.1007/JHEP07(2011)091}{{\em JHEP}
  {\bfseries 07} (2011) 091}, \href{http://arxiv.org/abs/1105.3911}{{\ttfamily
  arXiv:1105.3911 [hep-ph]}}.

\bibitem{JUNO:2015sjr}
{\bfseries JUNO} Collaboration, Z.~Djurcic {\em et~al.}, ``{JUNO Conceptual
  Design Report},'' \href{http://arxiv.org/abs/1508.07166}{{\ttfamily
  arXiv:1508.07166 [physics.ins-det]}}.

\bibitem{DUNE:2020lwj}
{\bfseries DUNE} Collaboration, B.~Abi {\em et~al.}, ``{Deep Underground
  Neutrino Experiment (DUNE), Far Detector Technical Design Report, Volume I
  Introduction to DUNE},''
  \href{http://dx.doi.org/10.1088/1748-0221/15/08/T08008}{{\em JINST}
  {\bfseries 15} no.~08, (2020) T08008},
  \href{http://arxiv.org/abs/2002.02967}{{\ttfamily arXiv:2002.02967
  [physics.ins-det]}}.

\bibitem{NOvA:2021nfi}
{\bfseries NOvA} Collaboration, M.~A. Acero {\em et~al.}, ``{Improved
  measurement of neutrino oscillation parameters by the NOvA experiment},''
  \href{http://dx.doi.org/10.1103/PhysRevD.106.032004}{{\em Phys. Rev. D}
  {\bfseries 106} no.~3, (2022) 032004},
  \href{http://arxiv.org/abs/2108.08219}{{\ttfamily arXiv:2108.08219
  [hep-ex]}}.

\end{thebibliography}\endgroup

\end{document}